
\input phyzzx

\newif\ifKUNS \KUNStrue
\Pubnum={\ifdraft\undertext{\strut$\mib draft$}\cr\fi
         KUNS-\the\pubnum}
\newcount\YEAR
\YEAR=\number\year
\global\advance\YEAR by-1900
\pubnum={\number\YEAR-XX}

\pubnum={1201}
\date={May 28
,~1993}
\titlepage
\title{\bf
Deflationary Universe Scenario}
\author{Boris~ Spokoiny \dag}\footnote{\dag}{ Address after June 23,~1993:
 ~Landau Institute for Theoretical
Physics, Russian Academy of Sciences, 142432 Chernogolovka, Moscow region,
Russia.}

\address{Department of Physics,~Kyoto University,
         ~Kyoto 606-01,~Japan}

\abstract{We show that it is possible to realize an inflationary
scenario even without conversion of the false vacuum energy to radiation.
Such cosmological models have a deflationary stage in which $Ha^2$ is
decreasing and radiation produced by particle creation in an expanding
Universe becomes dominant.
The preceding inflationary stage ends since the inflaton potential becomes
steep.
False vacuum energy is finally (partly) converted to the inflaton kinetic
 energy
, the potential energy rapidly decreases and the Universe comes to
the deflationary stage with a scale factor $a(t) \propto t^{1/3}$.
Basic features and observational consequences of this scenario are indicated.
}

\REF\Star{A.A.Starobinsky,Phys.Lett.91B (1980)99;
\nextline A.D.Linde,Phys.Lett.108B(1982);129B (1983)177;
\nextline A.Albrecht and P.J.Steinhardt,Phys.Rev.Lett.48(1982)1220.
}

\REF\Guth{ A.H.Guth,Phys.Rev.D 23 (1981)  347;
\nextline D.La \& P.J. Steinhardt, Phys. Rev.Lett. 62 (1989) 376. }

\REF\Rat{B.Ratra \& P.J.E.Peebles, Phys.Rev.D37 (1988)3406.}

\REF\Spo{B.L.Spokoiny, Vistas     in Astronomy 37 (1993)481;
 ~~Stochastic       approach      to  a   non
 de Sitter
Inflation,
Kyoto  Univ.  Preprint   KUNS-1192,  ~May (1993).}

\REF\Fuj{Y.Fujii and T.Nishioka, Phys.Rev.D42 (1990)361;ibid. D45 (1992)2140;
Phys.Lett. 254B (1991)347.}

\REF\Dol{A.D.Dolgov,in {\it The Very Early Universe},  Proceedings   of  the
Nuffield   Workshop,  Cambridge,  UK, 1982, edited  by  G. W. Gibbons and
S.T.Siklos.}

\REF\Ste{R.Crittenden and P.J.Steinhardt, Phys.Lett.293B (1992)32.}

\REF\Luc{F.Lucchin \& S.Matarrese ,
{\it Phys.Rev.}D{\bf 32}(1985)1316;  Phys.Lett. 164B  (1985)  282.
}

\REF\Hal{J.Halliwell,{\it Phys.Lett.} B{\bf 185} (1987) 341.
}

\REF\Bar{J.D.Barrow, Phys.Lett.B187 (1987) 12.}









\chapter{Introduction}
It has become almost an inflationary paradigm that after the end of the
superluminal (exponential) expansion stage driven by false - vacuum energy
there is  conversion of this energy to radiation.
There are two standard ways to realize this
 conversion:  by  decay of the oscillations of the
inflaton field near the minimum of its potential in the new or chaotic
 inflationary scenarios [\Star]
or by collisions of the
bubbles of new vacuum in the old or extended inflation [\Guth].
However for the solution of the famous problems such as horizon or flatness
problems and others
such a conversion is not a necessary condition.
In fact what one needs is some production of radiation that will become
dominant and finally will drive the evolution of the Universe.
In this paper we investigate basic features of this possibility in detail.

We consider inflationary models in which an inflaton potential is almost
flat in some region,say at $\varphi \leq \varphi_*$ so as the Universe
expands (quasi)exponentially while the inflaton is in this region and
in the region $\varphi > \varphi_*$  the inflaton potential becomes steep
enough so as inflation ceases.
In distinct to the standard models we consider the case when the potential
just rapidly falls for large values of $\varphi$
and does not have a minimum of the potential.
To understand how the Universe could become a radiation dominated one
it is enough to recall
that in an expanding Universe there is particle production.
At the inflationary stage the energy density of the produced particles
is inflated out but in the post - inflationary epoch the situation could
be changed.
If the potential falls rapidly enough the energy density $\varepsilon$
 of the produced
particles which are then thermalized can become dominated.
Since for relativistic particles (and radiation)
$\varepsilon \propto a^{-4}$
the stage in which $H^2 a^4$
is decreasing is necessary for the above phenomenon to occur.
We call this stage a deflationary one .
(On the contrary at the inflationary stage this quantity is rapidly
 increasing due to slow variation of $H$.)

There is an essential difference between the deflationary realization of
the inflationary scenario and the standard one.
In the ordinary realization an inflationary stage only provides some (good)
 initial conditions
 for the following
it standard radiation dominated stage .
The inflaton becomes \lq\lq frozen"
 in the minimum of the potential and does not
participate in the  dynamics at the radiation dominated stage.
In principle this circumstance restricts greatly the possibility of
 comparison of
 different inflationary models from observations.
This standard construction  was greatly affected by the success of the
nucleosynthesis theory and very severe restrictions
which follow from
this theory for the dynamics of the Universe during the nucleosynthesis
epoch and for any \lq nonstandard' contributions to the energy-momentum
tensor at this time.
So any alternative model in which an inflation ends in a different way
should have some mechanism to pass through the nucleosynthesis test
if one does not want a new fine tuning.

On the contrary in the deflationary realization an inflaton
 \lq \lq survives''
and its dynamics may be still important at later stages of the evolution
of the Universe,even at the present time.
For example inflaton can give us
dark matter.
In this case we have a natural bias since perturbations of the scalar field
decay when the wavelength becomes less than the size of the horizon.
This results in a more homogeneous distribution of mass in comparison with
galaxy distribution (see [\Rat
]).
There is also a natural mechanism that provides very small inflaton energy
 density
during the nucleosynthesis time which still allows to have (relatively)
large inflaton energy density now.
The idea is rather simple.
If the inflaton potential is steep enough at the deflationary stage then
the dynamics of the Universe is driven by the inflaton kinetic energy term
and the potential energy term is rapidly decaying.
(We call such a stage a critical deflationary one).
After the moment of time when radiation becomes dominant the kinetic energy
term starts to fall rapidly, the potential energy term is very small since
it was falling at the preceding critical deflationary stage but it becomes
almost frozen (since time derivative of the inflaton is rapidly decaying).
So there is a big  \lq\lq quasipure''
 radiation dominated epoch in this model
at which the contribution of the scalar field into the energy-momentum
tensor is negligible.
The nucleosynthesis period is supposed to be inside this epoch.
When the (falling) radiation energy density becomes of the order of (frozen)
inflaton potential energy density the Universe comes to the
 \lq \lq combined" stage at which the contribution of the inflaton energy is
com
that of ordinary matter.

\chapter{Dynamics of the Universe with a deflationary stage}

A simple inflationary model is given by the system of equations
$$
H^2 = {1\over 3M^2}( {{\dot{\varphi}^2}\over 2} + V(\varphi)),
\eqno(1)
$$
$$
\ddot{\varphi} + 3H\dot{\varphi} + V' = 0,
\eqno(2)
$$
where in some region ,say $\varphi \leq \varphi_*$, the logarithm of the
potential is flat:\nextline $-M(\log{V})' \ll 1$ and in the other region
$\varphi \geq \varphi_*$ it is steep.
Here $M = M_{pl}/\sqrt{8\pi}$.
So while $\varphi \leq \varphi_*$ the dynamics of the scalar field is slow
and the Universe expands quasi-exponentially:
$a(t) \propto \exp{(\int H dt)}$ (in the simplest model).
In eq.(2) one takes into account only the second and the third terms
disregarding the first one.
When the scalar field reaches the region of the steep potential its dynamics
becomes rapid with respect to the Universe expansion.
In eq.(2) the first and the third terms become the most important ones,the
second term being a small friction.
In an ordinary inflationary scenario the potential $V(\varphi)$ has a minimum
and the scalar field rapidly rolls down to the bottom of the potential and
starts to oscillate near the minimum of the potential.
These oscillations generate ordinary matter.

In this paper we consider another possibility of the transition to the
radiation dominated stage  and
consider an inflaton potential that does not have a minimum and just falls
 at large
 $\varphi$.
Matter is generated by particle production at the deflationary stage.
In the standard scenario the  potential energy of the scalar field is
transformed into the energy of oscillations, some part of the energy being
lost due to friction.
The transition from the region of the flat logarithm of the potential to
the region of the steep logarithm of the potential is sharp in the sense
that the transition proceeds in a (small) region $\Delta \varphi \sim M  $.
In our case there could be two possibilities of this transition : a sharp
transition and a smooth one with the transitional region
$\Delta \varphi \gg M$.
A simple example of the potential with a sharp transition is
$$
V(\varphi)= {\exp{(-\lambda_1 \varphi/M)}
 \over {1+ \exp{(\lambda_2 \varphi/M)   } }},
\eqno(3)
$$
where $\lambda_1 \ll 1$ and $\lambda_2$ is large enough.
A general form of the potential with a smooth transition is
$$
V(\varphi)= V_0 \exp{(-\lambda(\varphi){\varphi \over M})},
\eqno(4)
$$
where $\lambda(\varphi)$ is a slowly varying function of $\varphi$,
satisfying the condition
\footnote{*}{Stochastic approach to the inflationary
Universe driven by the scalar field with such type of potential was
 developed in [\Spo].}
$$
M\lambda' \ll 1.
\eqno(5)
$$
A simple example of this type of potential is just a gaussian potential
\nextline
$
V(\varphi) = V_0 \exp{(-{\varphi^2 \over 2f^2})}
$
with $ f^2 / M^2 \gg 1 $.

1)Sharp transition.
First we consider the case that is closer to the standard one: sharp
 transition to the region of the steep potential.
In this case  potential energy is converted to the kinetic energy
of the inflaton:
$$
\dot{\varphi}^2 \sim V(\varphi_*),
 \eqno(6)
$$
and the Universe becomes driven by the kinetic energy term.
This kinetic energy term is decaying due to friction.
Assuming that the potential energy does not affect the inflaton dynamics
we obtain from eq.(2)
 $$
\dot{\varphi} = {c_{\varphi}\over a^3},
\eqno(7)
$$
 where $c_{\varphi}$ is a constant
and $c_{\varphi} ^2 /a_0 ^6 \sim V(\varphi_*) $.
Substituting (7) into (1) we get
$$
a(t) \propto t^{1/3},
\eqno(8)
$$
$$
\varphi = \varphi_0  +  \sqrt{6}M \log{a}.
\eqno(9)
$$
Using (9) we find the condition for the decay of the potential:
$$
\epsilon(\varphi) = {\tilde{V}(\varphi) \over \tilde{V}(\varphi_*) } \ll 1,~~
where~~ \tilde{V}(\varphi) = V(\varphi)\exp{({\sqrt{6}\varphi \over M})}.
\eqno(10)
$$
 Thus if (10) is fulfilled the potential decays more rapidly
 than the inflaton
kinetic
energy term.

In an expanding Universe there is particle production.
The energy density of particles produced by gravitational field  for the
characteristic time $H^{-1}$   may be estimated as
 $\varepsilon \sim    10^{-3}nH^4 $ for the
scalar particles non-conformally coupled to gravity.
Here $n$ is the number of species and it is supposed that the mass of
 produced particles $m \ll H$.
The energy density of the particles
produced at the inflationary stage is inflated out
but the energy density of the above scalar particles produced
at the deflationary stage (where $Ha^2$ decreases)
$\varepsilon \propto {1 \over a^4}$ and falls less rapidly than the main
contribution (of the scalar field ) to the energy-momentum tensor.
Although initially the energy density of the produced particles is negligible
with respect to the main contribution due to the inflaton it becomes
 dominating after some time.
To investigate  dynamics of the Universe we are to add matter energy
density
$
 \varepsilon = c_0 a^{-4}
$
to the  inflaton energy density in eq.(1):
$$
({\dot{a}\over a})^2 = {1 \over 3M^2}({c_{\varphi}^2 \over a^6 } +
{c_0 \over a^4}).
\eqno(11)
$$
While
$
a \ll a_0 = {c_{\varphi}\over \sqrt{c_0}}
$
the inflaton contribution dominates and the Universe expands according to
 (8).
However at $a \sim a_0$ the contribution of radiation becomes of the same
order and at $a \gg a_0$ the Universe is radiation dominated.
The value of the Hubble parameter $H_{eq}$  at the moment when
 the contributions of
both types of matter become equal to each other (at $a = a_0$) gives us a
characteristic time of the change of epochs.
At the inflationary stage radiation energy density decays faster than
that of inflaton but at the inflaton kinetic energy dominated stage (8) the
situation is inverse .
This is why we call this stage a deflationary one.
Integration of eq.(11) gives us
$$
{t \over t_0} = {a \over a_0} \sqrt{1 + ({a \over a_0})^2} -
arcsinh {a \over a_0},
\eqno(12)
$$
where
$
t_0 = 2\sqrt{2}H_{eq} ^{-1} = 2\sqrt{3}M {c_{\varphi}^2 \over c_0 ^{3/2}}.
$
At $t \ll t_0$ we have the  deflationary stage
$$
a(t) \simeq a_0 (3t/t_0)^{1/3},
\eqno(13)
$$
at $t \gg t_0 $ - radiation dominated stage
$$
a(t) \simeq a_0 (t/t_0)^{1/2}.
\eqno(14)
$$

The length of the deflationary epoch may be estimated as follows.
At the moment $t_r$ when the contributions of matter and the scalar field
become equal to each other we have
$ 3M^2 H_* ^2 ({a_* \over a_r})^6  \sim 10^{-3} n H_* ^4 ({a_* \over a_r})^4,
$
where $H_*$ and $a_*$ are the values of the Hubble parameter and the scale
factor at the beginning of the deflationary stage and $a_r$ is the value of
the scale factor at $t=t_r$ (at the beginning of the radiation dominated
stage (14)).
Thus we obtain
$$
{a_r \over a_*} \sim 50 n^{-1/2} {M \over H_*}.
\eqno(15)
$$

The created particles
are thermalized and we may estimate the temperature $T_{eq}$ just after the
achievement of thermal equilibrium as follows.
Assuming that the created scalar particles are Higgs bosons
and interactions between them are mainly due to gauge bosons we may
estimate the interaction rate as $\Gamma \sim \alpha^2 T_{eq}$.
Thermal equilibrium is achieved when $\Gamma \sim H(t_{eq})$ .
We also have  relations
$ {\pi \over 30} g_* T_{eq} ^4  = 10^{-3} n H_* ^4 ({a_* \over a_{eq
}} )^4 $ and
$H(t_{eq}) = H_* ({a_* \over a_{eq}})^3  $
at the deflationary stage.
As a result we get
$$
T_{eq} \sim \alpha ({n \over 100 g_*})^{3 \over 8} H_*.
\eqno(16)
$$
Taking as usual $\alpha^2 \sim 10^{-3}$, $g_* \sim 10^2$, $n \sim 10$
we obtain $T_{eq} \sim {1 \over 300}H_*$.
As follows from the constraints on the amplitude of gravitational waves
which follow from the small anisotropy of the microwave background
$H_* < (10^{-4} \div 10^{-5})M_{pl}$.
Constraints from adiabatic perturbations typically give a lower value of
$H_*$.
Thus $T_{eq} <3 \times (10^{11} \div 10^{12}) Gev$ and is rather low.
We find from (15) that at the beginning of the radiation dominated stage
$$
T_r \sim ({n^3 \over 3g_*})^{1/4} 10^{-2}{H_* ^2 \over M^2}M
\sim 10^{-2} {H_* ^2 \over M^2}M .
\eqno(17)
$$
Thus in the above estimates $T_r < 0.5 (10^8 \div 10^{10}) Gev $.
On the other hand we need $T_r \gg T_{NUC} \sim 1Mev$ for the successful
 nucleosynthesis
in any case, which means $H_* \gg 0.5 \times 10^9 Gev$ and $T_{eq} \gg 2
 \times 10^6 Gev$.

At the radiation dominated stage (14)
 $~~ \dot{\varphi} \propto t^{-3/2} $ so the integral
$ \int dt \dot{\varphi} $ converges and the change of the value of the
scalar field for the whole radiation dominated epoch is
$ \Delta \varphi \sim M $.
This means that the potential $ V(\varphi) $ at the end of the radiation
dominated stage is of the same order of magnitude as that in the beginning
of this stage if we suppose that the potential has only one
 characteristic scale of variation - the (reduced) Planck mass $M$.
Thus the potential does not change essentially at the radiation
 dominated stage contrary to the preceding deflationary stage
where it
was falling rapidly according to (8)-(10).
Roughly speaking one may say that the potential is frozen at the radiation
dominated stage.
A similar analysis for a nonrelativistic matter dominated stage gives
qualitatively the same results.
This means that the above radiation dominated stage (or maybe already
 nonrelativistic matter stage) is not eternal and at
last the Universe will get
$$
 \varepsilon \sim V(\varphi) \sim V(\varphi_r),
\eqno(18)
$$
where $\varphi_r$ is the value of the scalar field at the beginning of
the radiation dominated stage,
and as follows from (2)
$$
\dot{\varphi}^2 \sim V(\varphi) \sim \varepsilon \propto 1/t^2
\eqno(19)
$$
Since we want for the combined stage (19) to begin after the nucleosynthesis
time we need for the potential to be steep enough:
$$
V(\varphi_r) < V(\varphi_*) \exp{(-\lambda_{eff}(\varphi_r - \varphi_*) )},
\eqno(20)
$$
where
$$
\lambda_{eff} = \sqrt{6} {\log{30 V(\varphi_*) \over \pi g_* T_{NUC} ^4}
\over \log{30 V(\varphi_*) \over \pi g_* T_r ^4}}.
 \eqno(21)
$$

Now we investigate the dynamics of the Universe with scalar field and matter.
So we should add a matter energy term into eq.(1):
$$
H^2 = {1\over 3M^2}(\varepsilon + {{\dot{\varphi}^2}\over 2} + V(\varphi)),
\eqno(22)
$$
Matter is supposed to satisfy the equation of state
$p=k\varepsilon $,
where
$ k = 1/3 $
for radiation and
$ k = 0 $
for dust.
Conservation of the energy-momentum tensor of the system \lq \lq scalar field
plus matter"
${T_k^i}_{;i} = 0 $
and the equation for the scalar field (2) give us
$
 \varepsilon = C_0 a^{-3(1+k)}.~
$
One may easily check that the transition solution from the matter
dominated stage value $ \varphi = \varphi_m $ to the combined
solution (19) is
$$
\varphi(t) \simeq \varphi_m + c t^{-{1-k \over 1+k}}
- {{1+k} \over {2(3+k)}}
{ V'(\varphi_r) } t^2 ,
\eqno(23)
$$
where $c$ is a constant, $k = 1/3$ for radiation and $k = 0$ for dust (non-
relativistic matter).
The second term in (23) describes the initially dominated but now falling
mode (7) and the third term describes the second mode of eq.(2) which is
 initially small but growing.
Solution (23) is valid at
$t \ll t_c = \sqrt{V(\varphi_m)/ V'^2 (\varphi_m) }$ .
It is easy to check that
$$
\varepsilon(t_c) \sim \lambda^2  V(\varphi_m)
 \sim \lambda^2  \dot{\varphi}^2 (t_c),
\eqno(24)
$$
where
$$
\lambda \equiv \lambda (\varphi) = M (\log{V})'.
\eqno(25)
$$
 So $t_c$ is a characteristic time after which the solution (19)
is achieved.
{}From eq.(24) we can make some qualitative conclusions about the behavior
of the Universe filled with matter and a scalar field.
If $\lambda \ll 1$ then the scalar field dominates.
This corresponds to the inflationary situation.
On the contrary if $\lambda \gg 1$ then matter dominates.
If $\lambda \sim 1$ then at least naively the contributions of matter and the
  scalar field are comparable.
Since we are not able to solve the system of the scalar field - metric
equations analytically for an arbitrary potential $V(\varphi)$,
to get some ideas about the behavior of the system in general case
  we consider
some exactly solvable types of the potential that  may be not the most
interesti
applications.
Let $\lambda$ be constant or a slowly varying function of $\varphi$
 satisfying
the condition (5).
This means that we consider potentials with a (pseudo)exponential asymptotic
 at $\varphi
\rightarrow \infty $ behavior (4).
Then we find from (19)
$$
\varphi \simeq \tilde{\varphi}_0 + A\log{t},~~~~~~~~~~A=2M/\lambda.
\eqno(26)
$$
For any equation of state the system of metric and scalar field equations
 has the solution (26)
\footnote{**}{Solution (26)-(29)  has been obtained in refs.
[\Rat,\Fuj].
It was  considered  in the \lq \lq model of a decaying
cosmological constant"
[\Fuj-\Dol] in [\Fuj] in the context different from ours.
In ref.[\Rat] cosmological consequences of a rolling homogeneous scalar
field were discussed without any relation to inflation.}
 with
$$
\tilde{\varphi}_0 = {A\over 2}\log{2(3p-1)M^2\over \lambda^2 V_0}
\eqno(27)
$$
and
$$
a(t) \propto t^p,
\eqno(28)
$$
where
$$
p = 2/3(1+k),~~~~~\varepsilon ={3p^2 M^2 \over t^2}(1 - 3(1 + k)/\lambda^2)
\eqno(29)
$$
The solution (26)-(29) exists if $ \lambda^2 \geq {\lambda_c}^2 = 3(1 + k)$,
the ratio of the inflaton energy density to the ordinary matter energy
density being
$$
{\varepsilon_{\varphi} \over \varepsilon}
= {3(1+k)/\lambda^2 \over (1 - 3(1+k)/\lambda^2)}.
\eqno(30)
$$
So we see that if the potential is steep enough inclusion of the scalar
field does not change the evolution behavior (28).
On the contrary if the potential is flat enough ($\lambda < \lambda_c $),
 the
 scalar field dominates
, matter is inflated out and we have the solution (28) with
$$
 p = 2/\lambda^2 ,~~~~~
\varphi= \varphi_0 ' + (2M/ \lambda ) \log{(Mt)},~~~~
 \varepsilon \propto t^{-2\lambda_c /\lambda} .
\eqno(31)
$$

2)Smooth transition.

Now we briefly consider dynamics in the case of a smooth transition to the
deflationary phase.
The inflaton potential is given by (4).

First let us consider for simplicity the case of a constant $ \lambda $.
In this case there is a  power-law solution of the system (1)-(2)
 $$
 a(t) \propto t^p,~~~~~
p = 2/\lambda^2 ,~~~~~
\varphi= \varphi_0  + (2M/ \lambda ) \log{(Mt)},
\eqno(32)
$$
If
$ \lambda (\varphi) \gsim 1 $
and (5) is fulfilled  we have an approximate
solution (32) where
$ \lambda = \lambda (\varphi) $.

If
$ \lambda(\varphi)<\sqrt{2}$ then $ p>1 $ and (32) describes an inflationary
stage.
If
$ \sqrt{2}< \lambda(\varphi) <2 $
then $ 1>p>1/2 $
and we will call this stage \lq\lq pseudo-inflation".
\footnote{***}{ Crittenden and Steinhardt [\Ste] recently considered
 inflation   with a Brans-Dicke type dilaton $\varphi$    and depending upon
 $\varphi$
parameter $\omega$.
The tunneling  of the inflaton to the true vacuum appears at the
 \lq\lq pseudo-inflationary" stage thus avoiding large bubbles.
In the Einstein frame the dilaton part of the CS model is equivalent
 to the model with the dilaton
potential (4).
In this paper we consider another possibility of the exit from the
inflationary stage without any tunneling.  }

If $ \lambda>2 $ then $ p<1/2 $ and we will call this stage deflation.
Solution (32) was obtained in [\Luc] ( for a constant $\lambda$)
 and considered mainly for an inflationary
stage in fact[8-10,4].
On the contrary our focus is
 mainly on the deflationary case $ p<1/2 $.
A more careful analysis shows that solution (32) is valid only if
$ \lambda<\sqrt{6} $ which corresponds to
$ p>1/3 $.
If $ \lambda > \sqrt{6} $ then solution (32) corresponds to negative $ V_0 $
in (4) and is not applicable.
In this case we should use an {\it asymptotic } at $ t \rightarrow \infty $
(critical deflationary) solution (8)-(9) and
$$
V(\varphi) \propto t^{-(2 + q)},
\eqno(33)
$$
where
$
q = 2({\lambda\over \sqrt{6}} - 1) > 0.
$
In the above solution it is supposed that
$
1 \gg V(\varphi)/\dot{\varphi}^2 \propto t^{-q}.
$

One may easily check that the energy density of particles produced in the
vicinity of the moment of time when $\ddot{a}=0~~$ is dominant.
Indeed the energy density produced at the moment $t_0 ~~ \rho(t) \propto
H^4 (t_0)/a^4 (t)$ and it is easy to see that the maximal production is
near the moment of time when $Ha$ is maximal.
This moment corresponds to $p=1$, where $p$ is the index of expansion
($a(t) \propto t^p$), and $\lambda(\varphi(t)) = \sqrt{2}$, thus just to
the end of the superluminal expansion.
The energy density of the produced particles is initially decreasing but
at the deflationary stage where $Ha^2$ is decreasing ($p < 1/2 $) it becomes
more and more dominant.
The deflationary stage starts when $\lambda(\varphi(t)) = 2 $.
Radiation produced after thermalization
of the particles created just after the end of the superluminal expansion
must become dominant in the Universe before its temperature drops below
 $T_{NUC}$.
This is a necessary condition for the model.

\centerline {Acknowledgments}

I am very grateful to Misao Sasaki for very useful discussions concerning
nucleosynthesis restrictions and growth of perturbations and to Ewan
Stewart and Jun'ichi Yokoyama for important stimulating comments which
 allowed to clarify
basic features of the model.
This work was supported in part by a JSPS  Fellowship and by
Monbusho Grant-in-Aid for Encouragement of Young Scientists, No. 92010.

\refout

\end